\documentstyle[seceq,epsf,twoside]{ptptex}
 
\newcommand{\bcen}{\begin{center}} 
\newcommand{\ecen}{\end{center}}

\newcommand{\sisi}{$\sigma\sigma$ } 
\newcommand{\pipi}{$\pi\pi$ } 
\newcommand{\kk}{$K\overline{K}$ }

\newcommand{\bc}{\begin{center}} 
\newcommand{\ec}{\end{center}}

\newcommand{\btablll}{\begin{tabular}{lll}} 
\newcommand{\btabllr}{\begin{tabular}{llr}} 
\newcommand{\btabll}{\begin{tabular}{ll}} 
\newcommand{\etab}{\end{tabular}} 
\notypesetlogo  
\title{${\mib{\pi-\pi}}$ interaction amplitudes with chiral constraints}
\author{Robert {\sc Kami\'nski}}
\inst{%
Henryk Niewodnicza\'nski Institute of Nuclear Physics,\\ 
PL 31-342 Krak\'ow, Poland
}
\abst{%
The \pipi interaction amplitudes have been calculated using a three coupled 
channel model both with and without constraints imposed by chiral models. 
Roy's equations have been used to compare the amplitudes and to study 
the role played by chiral constraints in the \pipi interaction.
}

\begin{document}
\maketitle

\setcounter{tocdepth}{4}

\section{Model}

The \pipi interactions are one of the main sources of information on
low energy physics. 
Their knowledge in scalar-isoscalar channel 
allows us to investigate scalar meson spectroscopy 
and to test the predictions of many theoretical models.

Amplitudes of the \pipi interactions should fulfill some basic
conditions like unitarity and crossing symmetry.
Described in 
\cite{kll2} 
three-coupled channel amplitudes for \pipi, \kk and \sisi interactions were constructed 
using a $S$-matrix unitary model without any constraints on model parameters.
Several soutions were obtained by fits to experimental data in effective two 
pion mass region $m_{\pi\pi}$ from the \pipi threshold to 1600 MeV.

Problem of the crossing symmetry 
for the \pipi scattering has been studied in the past using dispersion techniques
\cite{roy,basdevant} 
which led to the construction of Roy's equations.
Using them with parameters as presented in 
\cite{basdevant} 
we calculated new real parts (treated hereafter as output) 
for the \pipi amplitudes obtained in
\cite{kll2}.
The aim of our study was comparison of the output with real parts 
of given amplitudes (treated hereafter as input).
Roy's equations explicitly depend both on the energy behaviour of amplitudes and on
theresold parameters $a_{IJ}$ (for $I=0$ and 2)so such comparison was done for various \pipi threshold parameters 
in isoscalar and isotensor waves.

\section{Results}

\begin{figure}[h]
  \epsfxsize=12 cm
  \epsfysize=16 cm
 \centerline{\epsffile{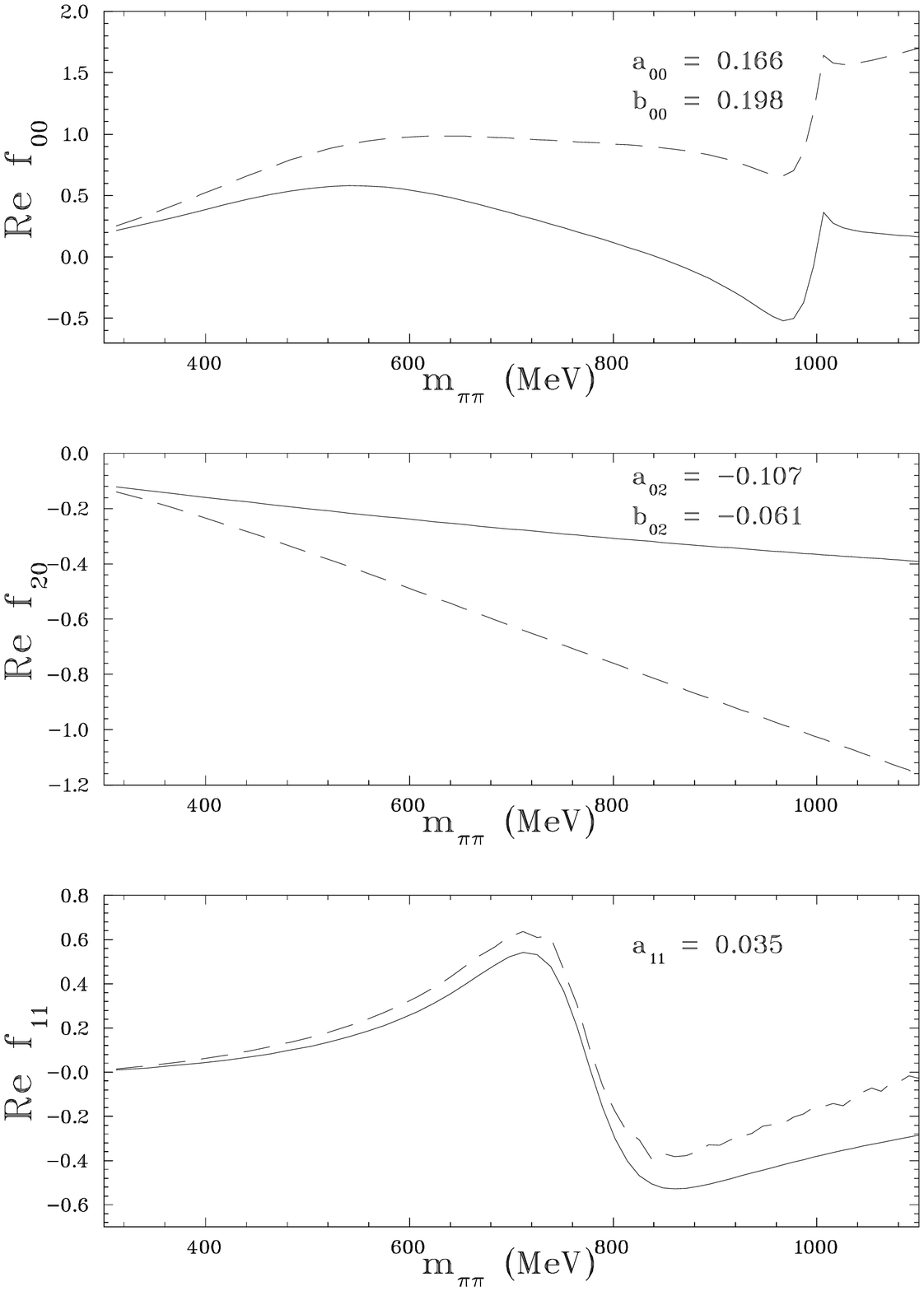}}
 \caption{Comparison of input (solid line) and output (dashed line) for the \pipi 
 amplitudes $f_{IJ}$ without any constraints for threshold parameters
 $a_{IJ}$ and $b_{IJ}$.}
  \label{fig:0}
\end{figure}

In Fig. 
\ref{fig:0}
one can see comparison of the output calculated using Roy's equations 
with input for solution A of 
\cite{kll2}.
 \begin{figure}[h]
  \epsfxsize=12 cm
  \epsfysize=16 cm
 \centerline{\epsffile{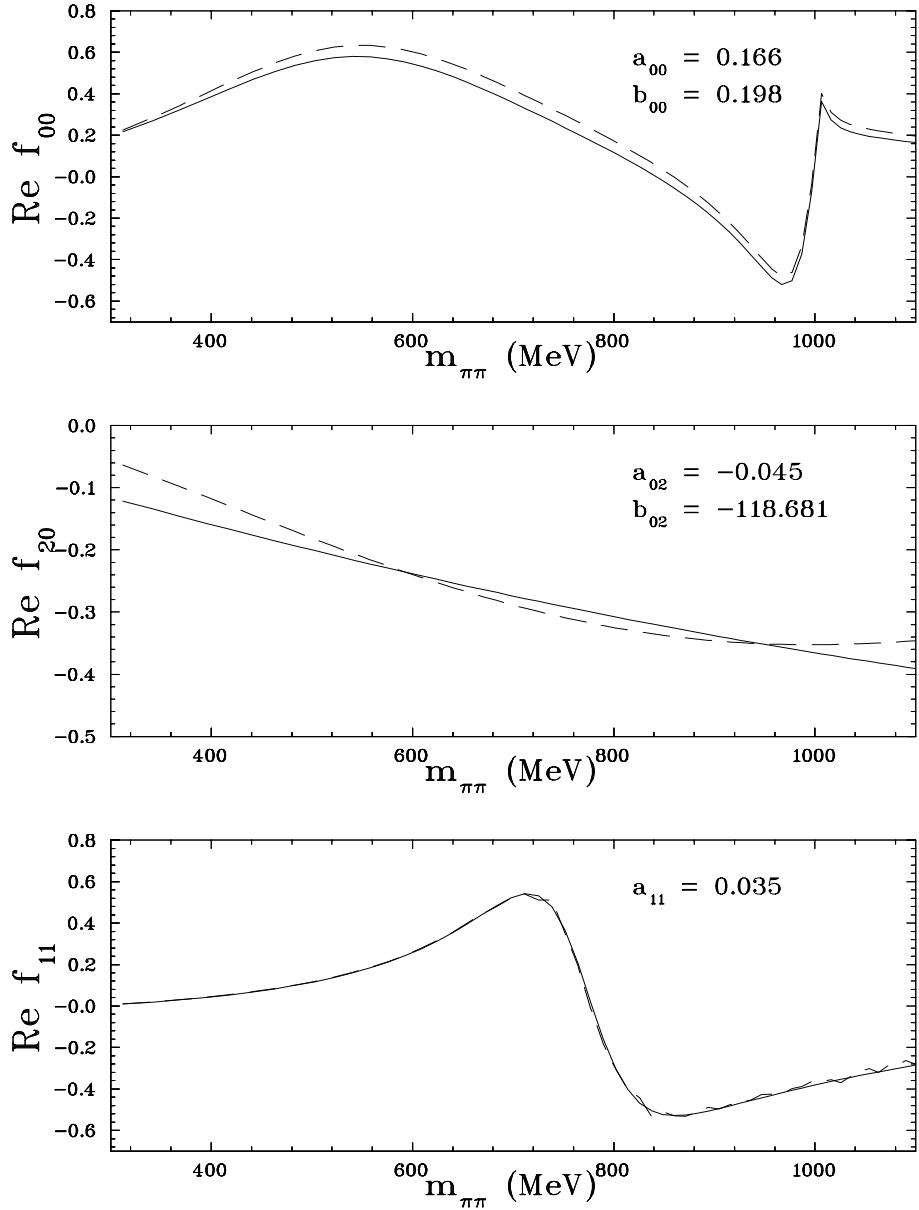}}
 \caption{Comparison of input (solid line) and output (dashed line) for the \pipi 
 amplitudes $f_{IJ}$ with only constraint for $a_{20}$}
  \label{fig:1}
\end{figure}
\begin{figure}[h]
  \epsfxsize=12 cm
  \epsfysize=16 cm
 \centerline{\epsffile{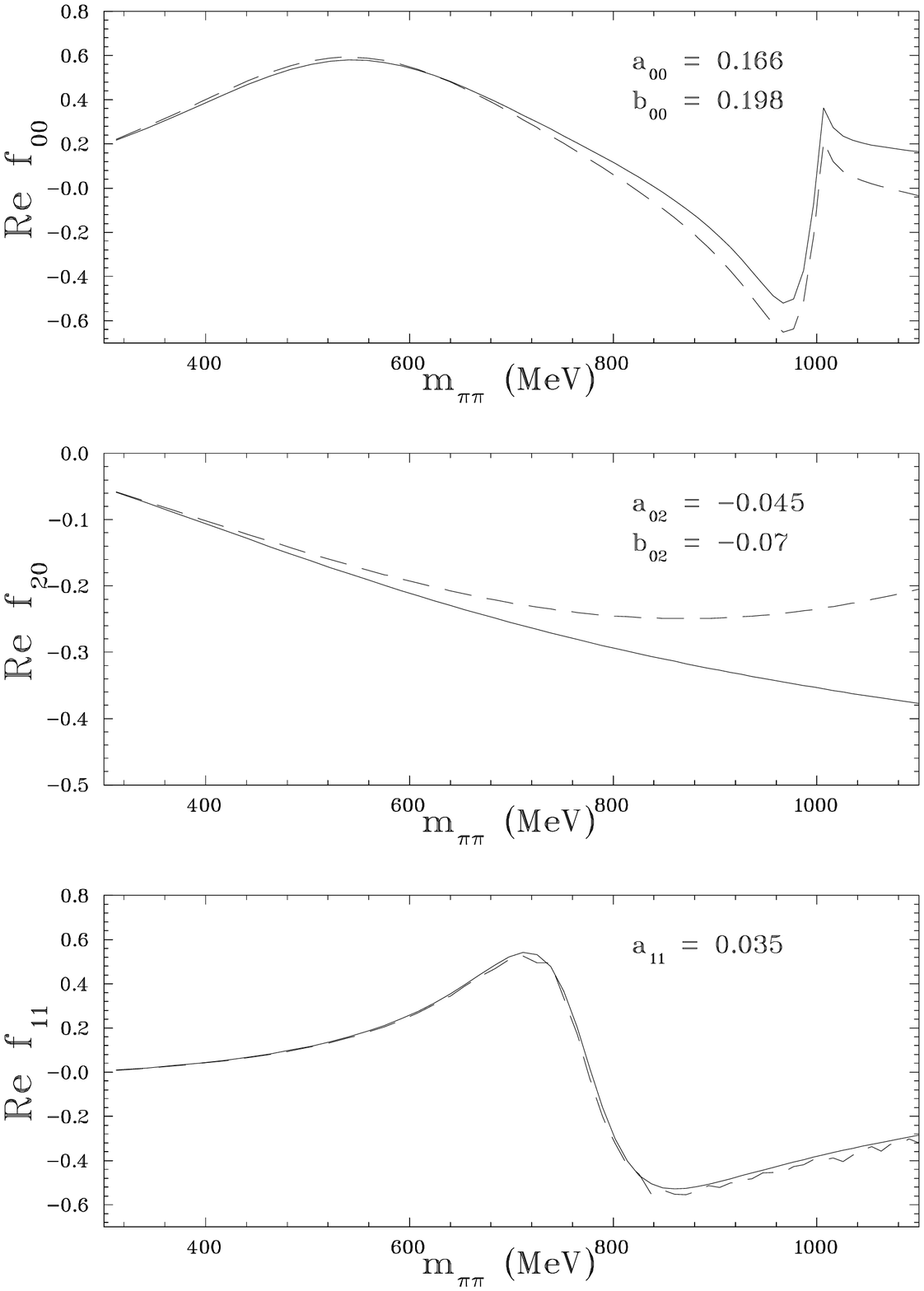}}
 \caption{Comparison of input (solid line) and output (dashed line) for the \pipi 
 amplitudes $f_{IJ}$ with constraints for $a_{20}$ and $b_{20}$.}
  \label{fig:2}
\end{figure}
\begin{figure}[h]
  \epsfxsize=12 cm
  \epsfysize=16 cm
 \centerline{\epsffile{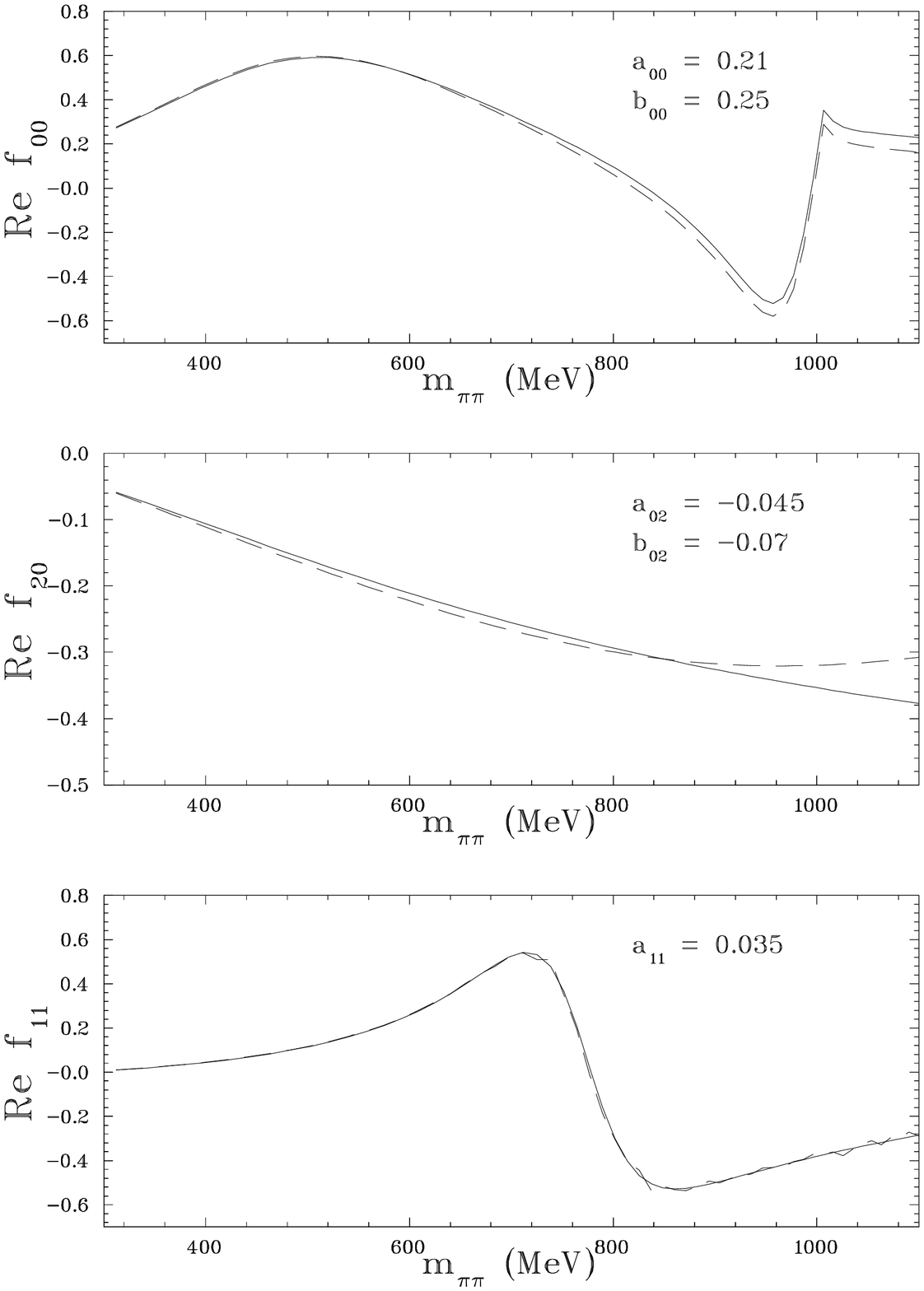}}
 \caption{Comparison of input (solid line) and output (dashed line) for the \pipi 
 amplitudes $f_{IJ}$ with constraints for $a_{20}$, $b_{20}$ and $a_{00}$.}
  \label{fig:3}
\end{figure}
The scattering lengths $a_{IJ}$ and the so called  slope parameters $b_{IJ}$ have been 
calculated using the threshold expansion:
\begin{eqnarray}
Re f_{IJ}(s) = (\frac{s-4}{4})^J \left[a_{IJ} + b_{IJ}(\frac{s-4}{4}) + ...\right],
\label{expansion}
\end{eqnarray}
where $I$ and $J$ denote isospin and total spin of the \pipi system, $s=(m_{\pi\pi}/m_{\pi})^2$.
The curves in Fig. 
\ref{fig:0}
correspond to the case when both $a_{IJ}$ and $b_{IJ}$ were treated as free parameters  
(not imposed by hand).
Significant differences between input and output can be seen for both isoscalar and
isotensor wave.

First correction was the change of 
the scattering length in the isotensor waves. 
Its value $a_{02} = -0.045$ follows roughly the results of two loop 
calculations of ChPT~\cite{bijnens}.
As can be seen in Fig.
\ref{fig:1}
such a constraint allows significant decrease 
of the differences between input and output 
in comparison with those seen in Fig. 
\ref{fig:0}.
The big differences that still exist near the \pipi threshold in 
isotensor wave indicated that the slope parameter $b_{20}$ had to 
be changed.
Value $b_{20}=-0.07$ was chosen according to two loop predictions of ChPT 
(see \cite{bijnens}).
As can be seen in Fig. 
\ref{fig:2} 
introduction of such a constraint for $b_{20}$ allows to obtain much better agreement 
between input and output near the \pipi 
threshold in the isotensor wave.

The last calculations were done for amplitudes with additional constraint for  
isoscalar scattering wave
$a_{00} = 0.21$.     
As can be seen in Fig. 
\ref{fig:3} 
inputs and outputs well agree in wide energy range up to 
about 1000 MeV
for both isoscalar and isotensor waves.

\section{Conclusions}

Additional constraints for isotensor and isoscalar
scattering lengths and slope parameter for the isotensor wave are necessary to obtain 
reasonably good agreement of the \pipi scattering amplitudes with those calculated 
using Roy's equations.
Numerical values of those parameters predicted by ChPT may be used as constraints.

\end{document}